\documentclass[11pt]{article}  
\usepackage{menuproc}
\usepackage{cite}
\usepackage{epsfig}
\usepackage{amsmath,amssymb}
\begin{document}
%
%
%
\titlematter{Progress in Meson-Nucleon Physics: Status and Perspectives}%
{Ulf-G.~Mei{\ss}ner}%
{Forschungszentrum J\"ulich, Institut f\"ur Kernphysik (Th), 
D-52425 J\"ulich, Germany}
{In this opening talk I will address some issues that exemplify the
theoretical progress that has been made and is to be expected in the field of
meson-nucleon physics. My emphasis will be on performing precision
calculations to test aspects of QCD, including also electroweak
probes. In addition, I discuss the problems and opportunities related to the
strange quark sector.}%
%

\section{Introduction: Precision and symmetries}

\begin{figure}[b]
\parbox{.2\textwidth}{\epsfig{file= 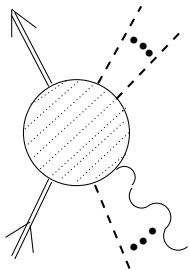,width=.15\textwidth,silent=,clip=}}
\hfill
\parbox{.75\textwidth}{\caption{\label{fig:facets}
A typical diagram showing the many facets of pion-nucleon (meson-baryon) physics. 
Here, solid, dashed and wiggly lines denote nucleons (baryons), pions
(Goldstone bosons) and photons (electroweak probes), in order. Pertinent
processes are $\pi N \to \pi N$, $\pi N \to \pi \pi N$, 
$\gamma^{(\star)} N \to\pi N$, $\gamma^{(\star)} N \to\pi \pi N$, 
$\gamma^{(\star)} N \to\gamma N$, $KN \to KN$, and many others. The physics encoded 
in these reactions covers chiral QCD dynamics, the structure of the nucleon and of resonances,
bound state dynamics, spin and polarization phenomena, electroweak interactions, and so on.
}}
\end{figure}

The field of meson--nucleon physics is a very rich one, a few typical examples
of processes to be discussed at this conference are shown in fig.~\ref{fig:facets},
also listed are some (but by far not all) of the pertinent physics issues. The aim
of this field is ambitious - one tries to understand QCD in the non--perturbative
regime where the strong coupling constant is large. Therefore, one also speaks of
{\em strong} {\em QCD}. This challenging theory can ultimately only be understood
through precise and systematic calculations matched by equally accurate data. I will
address here some of the theoretical developments that have been taken place over
the last years, concentrating on the use of symmetries. For low energy processes, 
we have a consistent calculational scheme based on the QCD symmetries and their 
realizations, chiral perturbation theory (CHPT). It is based on a systematic expansion
of S--matrix elements and transition currents in terms of small parameters. These
are external momenta and quark masses with respect to the typical hadronic scale of 
about 1 GeV. The relevant degrees of freedom are not quarks and gluons but rather
pions (Goldstone bosons) chirally coupled to nucleons (matter fields).
 As an example of a precise and systematic investigation I will consider
isospin violation in $\pi$N scattering in section~\ref{sec:iso}. 
Of course, CHPT does not allow to incorporate resonances
and bound states systematically, for that, one has to perform a non--perturbative
resummation. This can, however, be done in a fashion that preserves the low--energy structure
as demanded by CHPT, see section~\ref{sec:uni}, with applications to $\pi$N and $\bar K$N
scattering. In the last section, I mention some outstanding problems which require some
theoretical attention. Lastly, let me note that space forbids to discuss models, 
which can be quite useful or even indespensable, like e.g. in a systematic investigation
of the baryon spectrum. This is left to other speakers.

\section{Isospin violation in the pion-nucleon system}
\label{sec:iso}
\begin{figure}[tb]
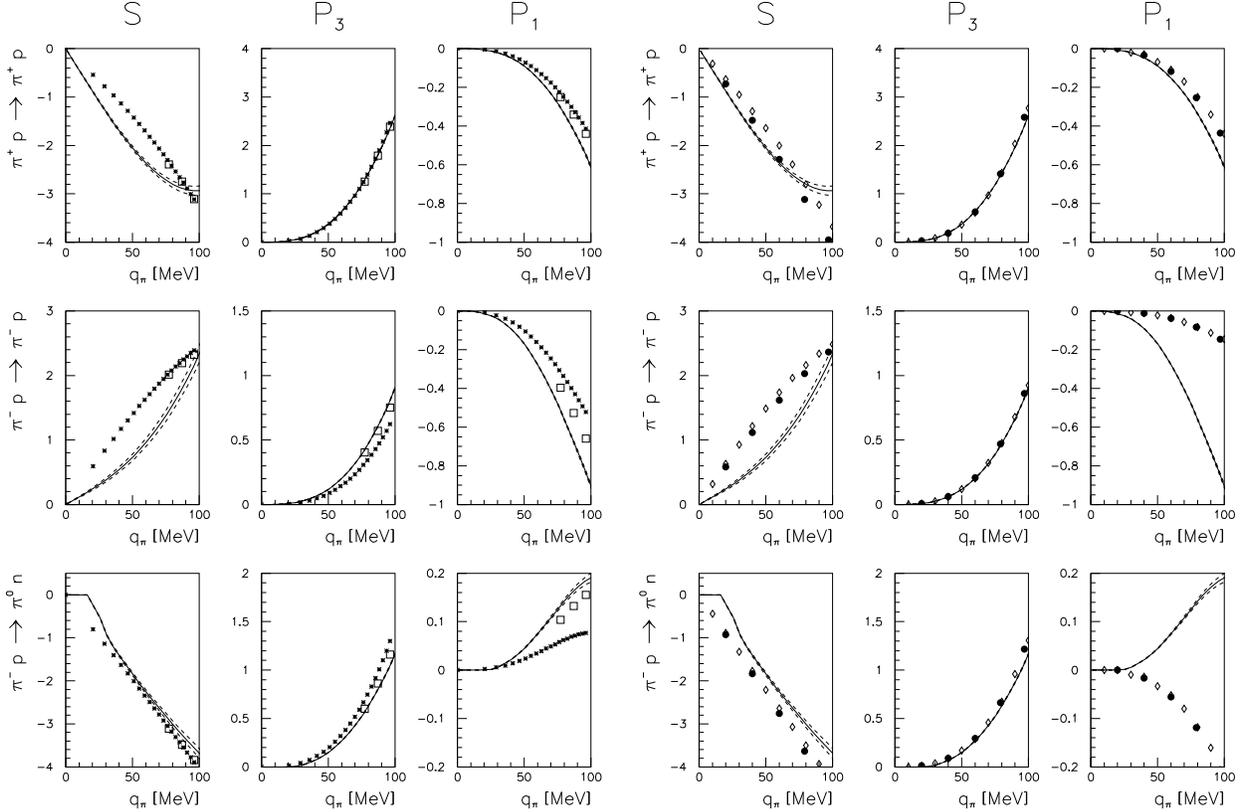

\parbox{.49\textwidth}{\epsfig{file= pinem.epsi,width=.48\textwidth,silent=,clip=}}
\hfill
\parbox{.49\textwidth}{\epsfig{file= pinks.epsi,width=.48\textwidth,silent=,clip=}}
\caption{\label{fig:pin}
Strong pion--nucleon phase shifts as a function of the 
pion laboratory momentum $q_\pi$ for the three measured channels.
Shown are the S--wave and the $j=1/2, \, 3/2$ P--waves. 
The solid line corresponds to the CHPT solution \cite{FM}, the dashed one 
to the one--sigma uncertainty range.
Left panel: Comparison to the 
EM98~\cite{mats} (stars) and the EM00~\cite{matsinew} (open squares) phases. 
Right panel:  Comparison to the
KA85~\cite{ka85} (full dots) and the SP98~\cite{sp98} (open diamonds) phases.}
\end{figure}

We now want to apply CHPT to one of the most studied processes, elastic
pion--nucleon scattering. More precisely, we will
consider systematically effects of isospin violation (I\!\!/) due to the
light quark mass difference, $m_u \neq m_d$, and electromagnetism, $q_u \neq q_d$. 
Before discussing in
some detail isospin violation in $\pi$N scattering, a few general
remarks are in order. In QCD plus QED, we have {\em two} sources of isospin
violation. In QCD, the light quark mass difference leads to isovector
terms, as reflected in the quark mass term (for two flavors)
\begin{equation}
{\cal H}_{\rm QCD}^{\rm mass} = m_u \bar u u + m_d \bar d d =
\frac{1}{2} (m_u+m_d) (\bar u u + \bar d d ) +
\frac{1}{2} (m_u-m_d) (\bar u u - \bar d d ) ~,
\end{equation}
where the last term on the right hand side is clearly of isovector
nature leading to strong I\!\!/. Naively, one could expect huge
I\!\!/ effects since $|(m_u-m_d)/(m_u+m_d)| \simeq 1/3$. However, the
scale one should compare to is the hadronic one, so that one indeed anticipates
very small effects, $(m_u -m_d)/ \Lambda_\chi < 1\%$. Only in
processes involving neutral pions one can expect much bigger effects \cite{neutralpi}.
The other source of I\!\!/ is
electromagnetism (em). Hadron mass shifts due to virtual photon
exchange between quarks can be estimated as $\delta m \simeq
\alpha_{\rm em} \cdot \Lambda_{\rm QCD} \cdot {\cal O}(1) \sim {\rm few} \,
{\rm MeV}$. In fact, typical electromagnetic mass splittings in 
meson and baryon multiplets are of this order. Therefore, these
two types of I\!\!/ have to be considered {\em consistently}. This can be
done by including virtual photons in the chiral effective Lagrangian
of pions and nucleons, treating the electric charge $e$ as another
small parameter. The machinery to do such calculations has been 
developed over the last years \cite{machinery}. To get a better idea about the size
of the I\!\!/ in $\pi$N scattering, let us a perform a lowest order
tree level analysis comparing elastic $\pi^- \pi^+$, $\pi^- K^+$
and $\pi^- p$ scattering. The first two processes can be taken from
the literature \cite{KU,KM}, 
\begin{eqnarray}
a(\pi^- \pi^+ \to \pi^- \pi^+) &=& \frac{M_{\pi^\pm}^2}{16 \pi F_\pi^2} 
\left\{ 1 + \frac{M^2_{\pi^\pm} - M^2_{\pi^0}}{M^2_{\pi^\pm}} \right\}
= a^{(\rm LO)}_{\pi\pi} \left\{ 1 + 0.064 \right\} ~, \nonumber \\
a(\pi^- K^+ \to \pi^- K^+) &=& \frac{M_{\pi^\pm}M_{K^\pm}}{8 \pi
  F_\pi^2 (M_{\pi^\pm} + M_{K^\pm}) } 
\left\{ 1 + \frac{M^2_{\pi^\pm} - M^2_{\pi^0}}{M_{\pi^\pm} M_{K^\pm}} \right\}
= a^{(\rm LO)}_{\pi K} \left\{ 1 + 0.018 \right\} ~,  
\end{eqnarray}
where $a^{(\rm LO)}$ denotes the leading order isosymmetric S-wave
scattering length. Note that the different normalization of the
$\pi\pi$ and $\pi$K scattering amplitudes has historic roots.
The relative suppression in the kaon case is due to the mass factor
$M_{\pi} / M_K \simeq 0.28$.  Therefore, in complete analogy one gets for
the pion--nucleon case
\begin{equation}\label{pines}
a(\pi^- p \to \pi^- p) =  \frac{M_{\pi^\pm}m_{p}}{8 \pi
  F_\pi^2 (M_{\pi^\pm} + m_p)} \left\{ 1 + 0.018
  \frac{M_{K^\pm}}{m_p} \right\} =
a^{(\rm LO)}_{\pi p} \left\{ 1 + 0.01 \right\}~,
\end{equation}
so that we can expect I\!\!/ effects of the order of one
percent. Before considering that system, I 
point out again that it is mandatory to consider {\em all}
possible I\!\!/ effects at a given order, i.e. that one can
get very misleading results if one considers only one particular
``dominant'' effect. A nice example 
are the QCD contributions to I\!\!/ in $\pi^+ K^- \to \pi^0 K^0$ \cite{KM,NT}, which are
of relevance for the $\pi K$ atom lifetime measurements at CERN \cite{dirac},
\begin{eqnarray}
a^{{\rm strong} {I\!\!/}} (\pi^+ K^- \to \pi^0 K^0) \propto
\frac{\epsilon}{\sqrt{3}} \,\,=\,\,
\frac{\epsilon}{\sqrt{3}} \,\Biggl\{ 
\underbrace{1 -\frac{M_K}{M_\pi}}_{\rm kinematical}
\,+\,\underbrace{\frac{M_K^2+M_\pi^2}{2M_KM_\pi}}_{\pi^0\eta-{\rm mixing}}
\,+\,\underbrace{\frac{M_K^2-M_\pi^2}{2M_KM_\pi}}_{\rm quark\,\,mass} \Biggr\} ~
,
\end{eqnarray}
having  distinguished ``kinematical'' effects
(due to meson mass splittings), ``$\pi^0 \eta$ mixing'' effects 
which modify the isospin symmetric amplitude by
factors of $\sin\epsilon$ or $\cos\epsilon$, with $\epsilon$ the
standard mixing angle $\sim \arctan[(m_d-m_u)/(m_s - \hat{m})]$, and
``quark mass insertions'' for the four--meson vertex.
It is obvious from the above that for the strong isospin violating contributions,
individual ``effects'' are much larger (and can even be of opposite
sign) than the total sum.
Thus, for a reliable determination of the size of isospin breaking in 
the strong interactions,  it is primordial to describe electromagnetic
and strong contributions consistently and to include all possible effects
to the order one is working.  This was achieved for the case of
pion--nucleon scattering in  the framework of chiral perturbation
theory to third order in ref.\cite{FM}, leading to a new phase shift
analysis (for pion lab momenta below 100\ MeV as deduced from the isospin symmetric fourth
order calculation~\cite{FM4}). The resulting S- and P-wave phases for
the three measured physical channels $\pi^\pm p \to \pi^\pm p$ and $\pi^-
p \to \pi^0 n$ (charge exchange) are shown in fig.~\ref{fig:pin}.
CHPT does not leave any doubt about the correct 
definition of the {\em hadronic} masses of pions and nucleons
(which are not the same for the pions as well as for the nucleons as often assumed), and allows 
to extract the strong part of the scattering amplitude in a unique
way. At this order, there is only one strong I\!\!/ violating operator
whose strength can be fixed from the $np$ mass difference. The em
corrections are a bit more subtle. First, there are one-- and
two--photon exchanges, the latter amount to a few percent correction
for the  kinematics pertinent to the existing data. More precisely, for
pion lab momenta $\vec{q}_\pi$, two--photon exchange is suppressed compared to one--photon
exchange by a factor $e^2 M_\pi / (32 |\vec{q}_\pi|) \leq 0.04$ for $|\vec{q}_\pi| \geq
10\,$MeV.
Then there are soft photon contributions in
terms of loops and external leg radiation. Only the sum of these is
IR finite and their contribution depends of course on the detector
resolution. We have used $\Delta E_\gamma = 10\,$MeV. In addition,
there are hard photon contributions encoded in contact terms with
undetermined low energy constants (LECs).
After determining the unknown LECs by a fit to experimental data, one can
switch off all electromagnetic interactions and describe QCD with unequal  
up-- and down--quark masses and $e^2 = 0$. The so--determined strong phase shifts 
(mostly) agree with those of previous works~\cite{ka85,mats,sp98,matsinew} 
in the P--waves, but one finds 
a sizeably different behavior in the S--waves (in particular for
$\pi^- p$ elastic scattering), compare fig.~\ref{fig:pin}.  
This difference can be traced back to the  
inclusion (in CHPT) or omission (in other approaches) 
of {\em non-linear} photon--pion--nucleon
couplings, i.e. vertices of the type $\bar N N \pi \pi \gamma$. Such
vertices are a consequence of chiral symmetry and thus must be included.
 Of course, these results need to be checked further,
in particular, one also has to extract the pertinent scattering
lengths. In any case, it also should be investigated how such
non-linear couplings can be included in the often used  dispersion theoretical
approaches to em corrections \cite{nor}. Given the hadronic amplitudes
constructed in \cite{FM}, one can 
address the question of isospin violation by studying the usual triangle 
relation involving elastic $\pi^\pm p$ scattering and the charge exchange 
reaction (for a general discussion of such triangle ratios, see \cite{FMt}
and references therein). 
An important advantage of the CHPT calculation lies in the fact that one can  
easily separate dynamical from static isospin
breaking, the latter are due to hadron mass differences. Dynamical isospin breaking 
only occurs in the S--wave and is very small, $\sim 0.75 \%$, in
agreement with the estimate given in eq.(\ref{pines}).
Static effects do not increase the size of isospin violation in the 
S--wave significantly; by no means can one account for 
the reported 7~$\%$ isospin breaking~\cite{gibbs,matsi}. These are
presumably due to a mismatch between the models for the strong and the
em interactions used in these works. 
Note also that one finds large error bars on the parameter values
in the CHPT analysis. In order to improve 
this situation, one would like to fit to more experimental data. However,  
a third order CHPT calculation allows to describe scattering data 
for pion laboratory momenta not much higher than 100~MeV, a region where the data situation  
is not yet as good as one would hope. A fourth order calculation would certainly allow 
to fit to data higher in energy, but, on the other hand, would also introduce many more 
unknown coupling constants. Since isospin breaking effects are expected to be most  
prominent in the low energy region,  one might question the usefulness
of extending the analysis to full one--loop (fourth) order. Additional data  
for pion--nucleon scattering at very low energies would be very helpful in this respect.  
Also a combined fit to several reactions involving nucleons, pions, and photons, 
e.g.\ pion electro-- and photoproduction, as well as $\pi N \to \pi \pi N$, would help in  
pinning down the fundamental low--energy constants more precisely.

\section{Expanding the borders: Higher energies, resonances and all that}
\label{sec:uni}
Going to higher energies, one has to implement unitarity constraints
(imaginary parts become more important with increasing energy)
as well as coupled channel dynamics. In addition, resonances appear,
which might be genuine quark model states or be dynamically generated
by strong final state interactions. Furthermore, relativistic effects
become more important with increasing energies. Therefore, one needs a
non-perturbative resummation scheme since in a perturbative theory
like CHPT, one can never generate a bound state or a resonance. There
exist many such approaches, but it is possible and mandatory to link such a scheme tightly
to the chiral QCD dynamics. I follow here the approach pioneered by
Oller and Oset \cite{OO} for meson interactions and demonstrate how 
this can be improved and extended for pion--nucleon \cite{MO1} and $\bar K$N
scattering~\cite{MO2}. To be specific, let us consider $\pi$N
scattering. The starting point is the T--matrix for any partial wave,
which can be represented in closed form if one neglects for the moment
the crossed channel (left-hand) cuts (for  more explicit details, see
\cite{MO1})
\begin{equation}
T = \left[ \tilde{T} (W) + g(s) \right]^{-1}~,
\end{equation}
with $W = \sqrt{s}$ the cm energy (as noted in \cite{MO1}, the analytical structure is much
simpler when using $W$ instead of $s$).
$\tilde{T}$ collects all local terms and poles (which can be
most easily  interpreted in the large $N_c$ world) and $g(s)$ is the
meson-baryon loop function (the fundamental bubble) that is resummed by e.g.
dispersion relations in a way to exactly recover the right-hand
(unitarity) cut contributions. The function $g(s)$ needs
regularization, this can be best done in terms of a subtracted
dispersion relation and using dimensional regularization (for details,
see \cite{MO1}). It is important to ensure that in the low-energy
region, the so constructed T--matrix agrees with the one of CHPT.
In addition, one has to recover the contributions from the left-hand
cut. This can be achieved by a hierarchy of matching conditions, e.g.
for the $\pi$N system one has
\begin{eqnarray}
{\cal O}(p) &:& \tilde{T}_1 (W) = T_1^{\chi} (W) ~, \quad 
{\cal O}(p^2) ~:~ \tilde{T}_1 (W) + \tilde{T}_2 (W) =  T_1^{\chi} (W)
+ T_2^{\chi} (W) ~, \nonumber \\
{\cal O}(p^3) &:& \tilde{T}_1 (W) + \tilde{T}_2 (W)+ \tilde{T}_3 (W) 
=  T_1^{\chi} (W)+ T_2^{\chi} (W) + T_3^{\chi} (W) +  \tilde{T}_1
(W) \, g(s) \,  \tilde{T}_1 (W)~,
\end{eqnarray}
and so on. Here, $T_n^{\chi}$ is the T--matrix calculated within
CHPT to ${\cal O}(q^n)$. Of course, one has to avoid double counting as
soon as one includes pion loops, this is achieved by the last term in
the third equation (loops only start at third order in this case).
In addition, one can also include resonance fields by saturating the
local contact terms in the effective Lagrangian through explicit meson and
baryon resonances (for details, see \cite{MO1}). In particular,
in this framework one can cleanly separate genuine quark resonances
from dynamically generated resonance--like states. The former require
the inclusion of an explicit field in the underlying Lagrangian, whereas
in the latter case the fit will arrange itself so that the couplings to 
such an explicit field will vanish (see e.g. the discussion of the $\rho$
as a genuine resonance versus the $\sigma$ as a dynamically generated state
in \cite{Olljap}). 
\begin{figure}[t]
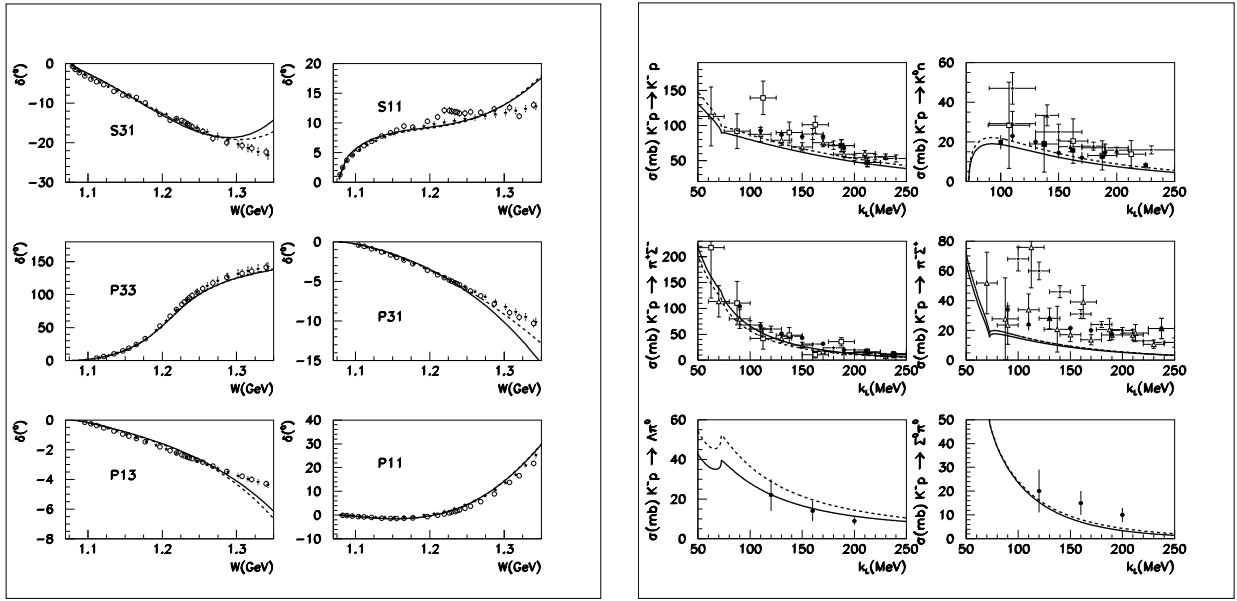

\parbox{.49\textwidth}{\epsfig{file= paw2.epsi,width=.48\textwidth,silent=,clip=}}
\hfill
\parbox{.49\textwidth}{\epsfig{file= scatf.epsi,width=.48\textwidth,silent=,clip=}}
\caption{\label{fig:fits}
Left panel: Fit to the low (S,P) $\pi$N partial waves. The solid (dashed)
lines refer to (un)constrained fits as explained in \cite{MO1}. Right
panel: Fit to various cross sections coupling to the $\bar K$N
channel. Solid lines: best fit, dashed lines: natural values for the
parameters, see \cite{MO2}.
}
\end{figure}
This method was
applied to $\pi$N scattering below the inelastic thresholds in \cite{MO1} by
matching to the third order heavy baryon CHPT results and including
the $\Delta (1232)$, $N^\star (1440)$, $\rho (770)$ and a scalar
resonance. Instead of the CHPT low--energy constants (LECs), one now fits
resonance parameters, of course, to a given order one can only determine
as many (combinations) thereof as there are LECs
A typical fit to the low partial waves is shown in the
left panel of fig.~\ref{fig:fits}. The threshold parameters are found
to be in good agreement with values obtained from phase shift analyses (for an updated
table, see e.g. \cite{Ollnstar}) and the $\Delta$ is found in the complex--W
plane at (1210-i53)~MeV, in good agreement with earlier findings
\cite{delta}. It is also important to point out that the
scalar exchange can be well represented by contact terms, i.e. no need
for a light sigma meson arises. 
These considerations were extended to
S--wave, strangeness $S = -1$ $\bar K$N scattering in \cite{MO2}. 
In this case, one has to consider the 
coupling to the whole set of SU(3) coupled channels, these are
$\bar K$N, $\Lambda \pi$, $\Sigma \pi$, $\Sigma \eta$ and $\Xi K$
(for earlier related work, see e.g. \cite{MuVal}). The lowest 
order (dimension one) effective Lagrangian was used, it depends on three
parameters, which are the average baryon octet mass and the pion decay
constant in the chiral limit and the subtraction constant appearing in
the dispersion relation for $g(s)$.
Their values can be estimated from simple considerations leading to the
so--called ``natural values''. 
One finds a good description of the
scattering data and the threshold ratios, see the dashed lines in the
right panel of fig.~\ref{fig:fits}. Leaving these parameters free, one
obtains the best fit (solid lines). It is worth to stress that the
values of the parameters of the best fit differ at most by 15\% from their natural
values. We have also investigated the pole structure of the S--wave
$\bar K$N system in the unphysical Riemann sheets. In addition to the
$I = 0$ pole close to the $\bar K$N threshold that can be identified
with the $\Lambda (1405)$ resonance, one finds another pole with $I =
0$ close to the $\Sigma \pi$ threshold and another one with $I = 1$
close to the $\bar K$N channel opening (which is threefold degenerate
in the isospin limit). Thus one can speculate about a nonet of $ J =
1/2$ meson--baryon resonances with strangeness $S = -1$. Still, one has
to investigate the $I = 1/2$ channel with $S =0, -2$ in this energy
interval to strengthen this conjecture. Also, one should include the
next--to--leading order terms and constrain the fit by $\pi N$ data
in this energy region (for related studies, see e.g. \cite{lu,nk}).
Note also that one can use
the results of ref.~\cite{MO2} to study
the S-wave $\Lambda \pi$ phase shift at the $\Xi$ mass which is
of relevance for CP violation studies in the decay $\Xi \to \Lambda
\pi \to p \pi \pi$~\cite{MO3}, see also \cite{tand} 
(for first experimental results, see \cite{luk}). Finally, such
methods can also be used to gain a better understanding of hadronic
final state interactions as it is mandatory to unravel aspects of
CP violation in B decays, see \cite{sue}.

\section{Summary and outlook}

Here, I will address some open issues which should be at the center
of theoretical investigations in the (near) future (or are already being
looked at). This list is, of course, highly subjective but I try to
cover as much ground as possible. Consider first the sector of 
the {\em non-strange} light quarks: 

\begin{itemize}
\item[$\bullet$] In the threshold region a precision machinery exists, which
  allows in particular to investigate the dual effects of strong and
  electromagnetic isospin violation. Much more work is needed to
  really pin down these subtle effects in various reactions.  Neutral
  pion scattering off protons should be measured, either via
  photoproduction \cite{Aron} or multiple scattering in $pp$
  collisions \cite{sven}.
\item[$\bullet$] The results of ref.~\cite{FM} for the em corrections to
  pion-nucleon scattering differ drastically from what has been
  available so far, see e.g. \cite{nor} or  the recent work by the ZuAC
  Collaboration \cite{zuac}. This deserves further study.
\item[$\bullet$] To incorporate also data from higher energies, as particularly
  stressed by H\"ohler, the dispersion relation machinery should be 
  married with chiral constraints, like e.g. using Roy-Steiner like
  equations, see e.g. \cite{BLP}, which has been proven fruitful in the 
  analysis of $\pi\pi$ and $\pi$K scattering.
\item[$\bullet$] Clearly, the bound state effective field theory calculations
 for pionic hydrogen and deuterium have to be finished/started to 
 deduce the precise S--wave scattering lengths from the accurate PSI
 data, see Rusetsky's talk at this Conference. 
\item[$\bullet$] The status of sigma term is still unsatisfactory. The present
  value say from the GWU group (see Pavan's talk) is uncomfortably
  large. A fresh look at strangeness in the proton might be useful.
\item[$\bullet$] Accurate photo/electroproduction  data have shed much light on
  the chiral dynamics of QCD, see the talks by Beck and
  Merkel. However, the new neutral pion  electroproduction data from MAMI-II 
  (at photon virtuality $Q^2 = 0.05\,$GeV$^2$) pose
   a serious challenge to theory. This needs to be resolved fast.
\item[$\bullet$] The inclusion of resonances is still an open problem with
 the exception of the delta, which can be included systematically if 
 one counts the nucleon-delta mass splitting as an additional free
 parameter (see the pioneering work in \cite{JM} and the
 systematization in \cite{HHK}).
\end{itemize}
I conclude that we are testing various aspects of strong QCD and need to
sharpen these investigations. In the {\em strange} quark sector, 
we are facing even more open problems, from which I mention a few:
\begin{itemize}
\item[$\bullet$] Clearly, more calculations based on the chiral unitary
  approach are needed, in particular the extension to photo-nuclear
  reactions (for some first attempt, see \cite{mupo}) and matching to
  higher order CHPT amplitudes than done so far. Obviously, one has to
  determine more parameters than in SU(2), but this should not be
  considered a barrier but rather an opportunity to gain a better
  understanding of e.g. SU(3) flavor breaking.
\item[$\bullet$] More accurate kaon photo/electroproduction data
  are needed to further test chiral baryon dynamics. A first analysis of the
  pioneering SAPHIR data from Bonn \cite{saphir} on kaon
  photoproduction off protons showed that this is feasible \cite{MSK}.
\item[$\bullet$] Such  production studies will also shed more light of the
 questions surrounding the nature of states like the $\Lambda (1405)$.
 The bound-state versus resonance scenario might e.g. be settled by
 measuring transition form factors, see e.g. \cite{s11}.
\item[$\bullet$] Since $m_s \sim \Lambda_{\rm QCD}$, the question remains
 whether one should consider the strange quark light or heavy? This
 can be further studied by making use e.g. of heavy kaon CHPT, see
 \cite{heavyK}. This allows to establish true SU(2) results within SU(3).
\item[$\bullet$] Very interesting is the question concerning the flavor
  dependence of chiral symmetry breaking. More specifically, does
  QCD have a complicated phase structure with $F_\pi^2 \neq 0$ and
  $\langle 0| \bar q q|0 \rangle^{N_f = 2}$ large (as indicated by the recent
  $K_{e4}$ data from BNL E865~\cite{BNL}, 
  see \cite{CGL}) for two flavors but a small condensate
  for the SU(3) case, $\langle 0| \bar q q|0 \rangle^{N_f = 3} \simeq
  0$ (which would neatly explain OZI violation in the scalar sector,
  see \cite{orsay})?
\end{itemize}
In summary, it is important to perform systematic and precise
calculations using as much as possible symmetry principles and
quantum field theoretical methods. Combining these with dispersion
theory, we can expect further progress in the field provided
that there is also progress in the experimental situation
for many processes involving pions, nucleons, photons and so on. 

\acknowledgments{It is a pleasure to thank  Nadia Fettes, Jos\'e Antonio
Oller and Bastian Kubis for many enjoyable collaborations. Bill Briscoe,
Helmut Haberzettl and their staff deserve praise for a well organized meeting.
I gratefully acknowledge financial support by the GW Center for Nuclear Studies.}


\end{document}